\documentstyle[12pt]{article}
\def\Re{{\cal R \mskip-4mu \lower.1ex \hbox{\it e}\,}}
\def\Im{{\cal I \mskip-5mu \lower.1ex \hbox{\it m}\,}}

\def\etal{{\it et al.}}

\def\beq{\begin{equation}}
\def\eeq{\end{equation}}
\def\brbsg{B(b\to s\gamma)}
\def\bsg{b\to s\gamma}
\def\Zbb{Z\rightarrow b\ov b}
\def\epsb{\epsilon_b}
\def\eps1{\epsilon_1}
\def\eps2{\epsilon_2}
\def\eps3{\epsilon_3}
\def\sub#1{_{\lower.25ex\hbox{$\scriptstyle#1$}}}
\def\sul#1{_{\kern-.1em#1}}
\def\sll#1{_{\kern-.2em#1}}
\def\sbl#1{_{\kern-.1em\lower.25ex\hbox{$\scriptstyle#1$}}}
\def\ssb#1{_{\lower.25ex\hbox{$\scriptscriptstyle#1$}}}
\def\sbb#1{_{\lower.4ex\hbox{$\scriptstyle#1$}}}

\def\GeV{\,{\rm GeV}}

\def\JL{J. L. Lopez}
\def\DVN{D. V. Nanopoulos}

\def\to{\rightarrow}
\def\ov{\overline}
\def\mh{\ifmmode m\sbl H \else $m\sbl H$\fi}
\def\mch{\ifmmode m_{H^\pm} \else $m_{H^\pm}$\fi}
\def\mt{\ifmmode m_t\else $m_t$\fi}
\def\mc{\ifmmode m_c\else $m_c$\fi}
\def\mz{\ifmmode M_Z\else $M_Z$\fi}
\def\mw{\ifmmode M_W\else $M_W$\fi}
\def\mws{\ifmmode M_W^2 \else $M_W^2$\fi}
\def\mhs{\ifmmode m_H^2 \else $m_H^2$\fi}
\def\mzs{\ifmmode M_Z^2 \else $M_Z^2$\fi}
\def\mts{\ifmmode m_t^2 \else $m_t^2$\fi}
\def\mcs{\ifmmode m_c^2 \else $m_c^2$\fi}
\def\mchs{\ifmmode m_{H^\pm}^2 \else $m_{H^\pm}^2$\fi}
\def\ztwo{\ifmmode Z_2\else $Z_2$\fi}
\def\zone{\ifmmode Z_1\else $Z_1$\fi}
\def\mtwo{\ifmmode M_2\else $M_2$\fi}
\def\mone{\ifmmode M_1\else $M_1$\fi}
\def\tb{\ifmmode \tan\beta \else $\tan\beta$\fi}
\def\xw{\ifmmode x\sub w\else $x\sub w$\fi}
\def\ch{\ifmmode H^\pm \else $H^\pm$\fi}
\def\lum{\ifmmode {\cal L}\else ${\cal L}$\fi}
\def\inpb{\ifmmode {\rm pb}^{-1}\else ${\rm pb}^{-1}$\fi}
\def\infb{\ifmmode {\rm fb}^{-1}\else ${\rm fb}^{-1}$\fi}
\def\epem{\ifmmode e^+e^-\else $e^+e^-$\fi}
\def\ppb{\ifmmode \bar pp\else $\bar pp$\fi}

\newskip\zatskip \zatskip=0pt plus0pt minus0pt
\def\coeff#1#2{{\textstyle{#1\over #2}}}
\def\matth{\mathsurround=0pt}
\def\lsim{\mathrel{\mathpalette\atversim<}}
\def\gsim{\mathrel{\mathpalette\atversim>}}
\def\atversim#1#2{\lower0.7ex\vbox{\baselineskip\zatskip\lineskip\zatskip
  \lineskiplimit 0pt\ialign{$\matth#1\hfil##\hfil$\crcr#2\crcr\sim\crcr}}}

\renewcommand{\thefootnote}{\fnsymbol{footnote}}

\begin{document} \begin{titlepage}
\setcounter{page}{1}
\thispagestyle{empty}
\rightline{\vbox{\halign{&#\hfil\cr
&CTP-TAMU-69/93\cr
&October 1993\cr}}}
\vspace{0.1in}
\begin{center}
\vglue 0.3cm
{\Large\bf $b\rightarrow s\gamma$ and $\epsilon_b$ Constraints on\\}
\vspace{0.2cm}
{\Large\bf Two Higgs Doublet Model\\}
\vglue 1.5cm
{GYE T. PARK\\}
\vglue 0.4cm
{\em Center for Theoretical Physics, Department of Physics, Texas A\&M
University\\}
{\em College Station, TX 77843--4242, USA\\}
{\em and\\}
{\em Astroparticle Physics Group, Houston Advanced Research Center
(HARC)\\}
{\em The Woodlands, TX 77381, USA\\}
\baselineskip=12pt

\end{center}

\begin{abstract}

We perform a combined analysis of two stringent constraints
on the 2 Higgs doublet model, one coming from the recently announced  CLEO II
bound on $\brbsg$ and the other from the recent LEP data on $\epsb$.
We have included one-loop vertex corrections to $\Zbb$ through $\epsb$ in the
model.
We find that the new $\epsb$ constraint excludes most of the less appealing
window $\tan\beta\lsim 1$ at $90\%$C.~L. for $m_t=150\GeV$.
We also find that although $\bsg$ constraint is stronger for $\tan\beta>1$,
$\epsb$ constraint is stronger for $\tan\beta\lsim 1$, and therefore these two
are the strongest and complimentary
constraints present in the charged Higgs sector of the model.

\end{abstract}

\renewcommand{\thefootnote}{\arabic{footnote}} \end{titlepage}
\setcounter{page}{1}


Despite the remarkable successes of the Standard Model(SM) in its complete
agreement with current all experimental data, there is still no
experimental information on the nature of its Higgs sector.
The 2 Higgs doublet model(2HDM) is one of the mildest extensions of the SM,
which has been consistent with experimental data.
In the 2HDM to be considered here, the Higgs sector consists of 2 doublets,
$\phi_1$ and $\phi_2$,
 coupled to the charge -1/3 and +2/3 quarks, respectively, which will ensure
the absence of Flavor-Changing Yukawa couplings at the tree level
 \cite{NOFC}. The physical Higgs spectrum of the model includes two CP-even
 neutral Higgs($H^0$, $h^0$), one CP-odd neutral Higgs($A^0$)
, and a pair of charged Higgs($H^\pm$). In addition to the masses of these
Higgs, there is another free parameter in the model, which is $\tan\beta\equiv
v_2/v_1$, the ratio of the vacuum expectation values of both doublets.

With a renewed interest on the
flavor-changing-neutral-current (FCNC) $\bsg$ decay, spurred by the
CLEO bound $\brbsg<8.4\times10^{-4}$ at $90\%$ C.L. \cite{CLEO}, it was pointed
out recently that the CLEO bound can be violated due to the charged Higgs
contribution in the 2HDM and the Minimal Supersymmetric Standard Model(MSSM)
basically if $m_{H^\pm}$ is too light, excluding large portion of the charged
Higgs parameter space \cite{BargerH}.
The recently announced CLEO II bound $\brbsg<5.4\times10^{-4}$ at
$95\%$\cite{Thorndike} excludes even larger portion of the parameter space
\cite{VernonHARC}. It has certainly proven that this particular decay mode can
provide more stringent constraint on new physics beyond SM than any other
experiments\cite{bsgamma}.
In our previous work\cite{Rbbsg2HD}, we pointed out that in addition to the
constraint from $\bsg$, the recent LEP data on $R_b (\equiv
{\Gamma(Z\rightarrow b\ov b)\over{\Gamma(Z\rightarrow hadrons)}})$
\cite{LP93} provides a mild additional constraint to the 2HDM.
In this work, we will show that the recent LEP data on a new observable
$\epsilon_b$
provides much stronger constraint, excluding at 90\% C.L. most of the parameter
space $\tan\beta\lsim 1$, which is a less appealing
 window
simply due to the apparent mass hierarchy $m_t\gg m_b$.
$\epsb$ has been recently introduced by Altarelli et.~al.
\cite{ABC,Altlecture}
, who have proposed a new scheme analyzing precision electroweak tests where
four variables, $\epsilon_{1,2,3}$ and $\epsilon_b$
are defined in a model independent way. These four variables correspond to
a set of observables $\Gamma_{l}, \Gamma_{b}, A^{l}_{FB}$ and $M_W/M_Z$.
The advantage of using these variables is that one need not specify
$m_t$ and $m_H$.
Among these variables, $\epsb$ is the most interesting observable for one to
consider in the 2HDM although $\epsilon_1$ can also provide an important
constraint, in the MSSM\cite{BFC,ABC} and a class of supergravity
models\cite{bsgamma,ewcorr}, due to a significant negative shift coming from
light chargino loop in the Z wave function renormalization
with the chargino mass $\sim {1\over2} M_Z$.
In fact, Altarelli et.~al. have applied the new $\epsilon$-analysis to the
MSSM, and their conclusion is that the model is in at least as good an
agreement with the data as the SM\cite{ABCMSSM}. Here we intend to do a similar
analysis in the framework
of 2HDM.

In the 2HDM, $\bsg$ decay receives contributions from penguin diagrams with
$W^\pm-t$ loop and $H^\pm-t$ loop.
The expression used for $\brbsg$ is given by \cite{BG}
\beq
{B(b\to s\gamma)\over B(b\to ce\bar\nu)}={6\alpha\over\pi}
{\left[\eta^{16/23}A_\gamma
+\coeff{8}{3}(\eta^{14/23}-\eta^{16/23})A_g+C\right]^2\over
I(m_c/m_b)\left[1-\coeff{2}{3\pi}\alpha_s(m_b)f(m_c/m_b)\right]},
\eeq
where $\eta=\alpha_s(M_Z)/\alpha_s(m_b)$, $I$ is the phase-space factor
$I(x)=1-8x^2+8x^6-x^8-24x^4\ln x$, and $f(m_c/m_b)=2.41$ the QCD
correction factor for the semileptonic decay.
We use the 3-loop expressions for $\alpha_s$ and choose $\Lambda_{QCD}$ to
obtain $\alpha_s(M_Z)$ consistent with the recent measurements at LEP.
In our computations we have used: $\alpha_s(M_Z)=0.118$, $ B(b\to
ce\bar\nu)=10.7\%$, $m_b=4.8\GeV$, and
$m_c/m_b=0.3$. The $A_\gamma,A_g$ are the
coefficients of the effective $bs\gamma$ and $bsg$ penguin operators
evaluated at the scale $M_Z$. The contributions to $A_{\gamma ,g}$ from the
$W^\pm-t$ loop, the $H^\pm-t$ loop are given in Ref\cite{BG}.
As mentioned above, the CLEO II bound excludes a large portion of the parameter
space. In Fig. 1 we present the excluded regions in ($m_{H^\pm}$,
$\tan\beta$)-plane for $m_t=130$, and $150\GeV$, which lie to the left of each
curve (solid). We have also imposed in the figure the lower bound on
$\tan\beta$ from  ${m_t\over{600}}\lsim\tan\beta\lsim{600\over{m_b}}$ obtained
by demanding that the theory remain perturbative\cite{BargerLE}.
We see from the figure that at large $\tan\beta$ one can obtain a lower bound
on $m_{H^\pm}$ for each value of $m_t$. And we obtain the bounds
, $m_{H^\pm}\gsim 186, 244\GeV$ for $m_t=130, 150\GeV$, respectively.

Following Altarelli et.~al.\cite{ABC}, $\epsb$ is defined from $\Gamma_b$, the
inclusive partial width for $\Zbb$,

\begin{equation}
\Gamma_b=3 R_{QCD} {G_FM^3_Z\over 6\pi\sqrt 2}\left(
1+{\alpha\over 12\pi}\right)\left[ \beta _b{\left( 3-\beta
^2_b\right)\over 2}{g^b_V}^2+\beta^3_b {g^b_A}^2\right] \;,
\end{equation}
with
\begin{eqnarray}
R_{QCD} &\cong&\left[1+1.2{\alpha_S\left(
M_Z\right)\over\pi}-1.1{\left(\alpha_S\left(
M_Z\right)\over\pi\right)}^2-12.8{\left(\alpha_S\left(
M_Z\right)\over\pi\right)}^3\right] \;,\\
\beta_b&=&\sqrt {1-{4m_b^2\over M_Z^2}} \;, \\
g^b_A&=&-{1\over2}\left(1+{\epsilon_1\over2}\right)\left(
1+{\epsb}\right)\;,\\
{g^b_V\over{g^b_A}}&=&{1-{4\over3}{\ov s}^2_W+\epsb}\over{1+\epsb} \;.
\end{eqnarray}
where ${\ov s}^2_W$ is an effective $\sin^2\theta_W$ for on-shell Z and the
explicit expression for $\epsilon_1$ is given in Ref\cite{BFC,ewcorr}.
$\epsb$ is closely related to the real part of the vertex correction to $\Zbb$
, $\nabla_b$
defined in Ref\cite{BF}.
In the SM, the diagrams for $\nabla_b$  involve
top quarks and $W^\pm$ bosons\cite{RbSM}. However, in the 2HDM there are
additional
diagrams involving $H^\pm$ bosons instead of $W^\pm$ bosons. These additional
diagrams
have been calculated in Ref\cite{Rbbsg2HD,Rb2HD,BF,Denner}. The charged Higgs
contribution to $\nabla_b$ is given as \cite{BF}
\begin{equation}
\nabla_b^{H^\pm}={\alpha\over 4\pi \sin^2\theta_W}\left[
{2 v_L F_L+2 v_R F_R}\over {v_L^2+v_R^2}
\right] \;,
\end{equation}
where $F_{L,R}=F_{L,R}^{(a)}+F_{L,R}^{(b)}+F_{L,R}^{(c)}$ and
\begin{eqnarray}
F_{L,R}^{(a)} &=& b_1\left(M_{H^+}, m_t, m_b\right) v_{L,R}
\lambda^2_{L,R}\;,\\
F_{L,R}^{(b)} &=&\left[\left({M_Z^2\over{\mu^2}}
c_6\left(M_{H^+}, m_t, m_t\right)-{1\over 2}-c_0\left(M_{H^+}, m_t,
m_t\right)\right)v_{R,L}^t\right. \nonumber \\
&& \hspace*{1.05in} \left. +{m_t^2\over{\mu^2}}
c_2\left(M_{H^+}, m_t, m_t\right)v_{L,R}^t\right]\lambda^2_{L,R}\;,\\
F_{L,R}^{(c)} &=& c_0\left(m_t, M_{H^+}, M_{H^+}\right)\left({1\over 2}-
\sin^2\theta_W\right)\lambda^2_{L,R}\;,
\end{eqnarray}
where $\mu$ is the renormalization scale and
\begin{eqnarray}
v_L &=& -{1\over 2}+{1\over 3}\sin^2\theta_W\,,  \quad v_R={1\over
3}\sin^2\theta_W \;, \\
v_L^t &=& {1\over 2}-{2\over 3}\sin^2\theta_W\,,  \quad v_R^t=-{2\over
3}\sin^2\theta_W \;, \\
\lambda_L &=& {m_t\over{{\sqrt 2} M_W \tan\beta}}\,,
\quad\lambda_R = {m_b \tan\beta\over{{\sqrt 2} M_W }}\;.
\end{eqnarray}
The $b_1$ and $c_{0,2,6}$ above are the reduced Passarino-Veltman
functions\cite{BF,Ahn}.
The charged Higgs contribution to $\epsb$, which is negative, grows as
$m^2_t/\tan^{2}\beta$ for $\tan\beta\ll{m_t\over{m_b}}$ as is seen from
Eq.~(13).
In our calculation, we neglect  the neutral Higgs contributions to $\nabla_b $
which are all proportional to $m_b^2\tan^2\beta$ and
become sizable only for
$\tan\beta>{m_t\over{m_b}}$ and very light neutral Higgs $\lsim50\GeV$, but
decreases rapidly to get negligibly small as the Higgs masses become
$\gsim100\GeV$\cite{Denner}.
We also neglect oblique corrections from the Higgs bosons just to avoid
introducing more paramters. However, this correction can become sizable
when  there are large mass splittings between the charged and neutral Higgs,
for example, it can grow as $m^2_{H^\pm}$ if
$m_{H^\pm}\gg m_{H^0,h^0,A^0}$.
Although $\tan\beta\gg1$ seems more appealing because of apparent hierarchy
 $m_t\gg m_b$, there are still no convincing arguments against $\tan\beta<1$.
 Our goal here is to see if one can put a severe constraint in this region.
In Fig. 1 we also show  the contours (dotted) of a predicted value of
$\epsb=-0.00592$,
which is the LEP lower limit at $90\%$C.~L.\cite{ABC,Altlecture}.
The excluded regions lie below each dotted curve for given $m_t$.
We do not consider higher values of $m_t$ here because the SM prediction for
$\epsb$
exceeds the LEP value already for $m_t\gsim 163\GeV$ \cite{ABC}.
For $m_t=150(130)\GeV$, $\tan\beta\lsim 1.03(0.51)$ is ruled out at $90\%$C.~L.
for $m_{H^\pm}\lsim 400\GeV$, and $\tan\beta\lsim 0.69(0.34)$
for $m_{H^\pm}\lsim 800\GeV$. We note that these strong constraints for
$\tan\beta\lsim 1$
stem from large deviations of $\epsb$ from the SM prediction, which grows as
$m^2_t/\tan^{2}\beta$ as explained above.
We have also considered other constraints from low-energy data primarily in
$B-\ov{B}, D-\ov{D}, K-\ov{K}$ mixing that exclude low values of
$\tan\beta$\cite{BargerLE,LowEdata}. But it turns out that none of them can
hardly compete with the present $\epsb$ constraint\cite{assume1}.
Nevertheless, the CLEO II
 bound is still by far the strongest constraint present in the charged Higgs
sector of the model for $\tan\beta> 1$. Therefore, we find that $\bsg$ and
$\epsb$ serve as the presently strongest and complimentary constraints in 2HDM.

In conclusion, we have performed a combined analysis of two stringent
constraints
on the 2 Higgs doublet model, one coming from the recently announced  CLEO II
bound on $\brbsg$ and the other from the recent LEP data on $\epsb$.
We have included one-loop vertex corrections to $\Zbb$ through $\epsb$ in the
model.
We find that the new $\epsb$ constraint excludes most of the less appealing
window $\tan\beta\lsim 1$ at $90\%$C.~L for $m_t=150\GeV$.
We also find that although $\bsg$ constraint is stronger for $\tan\beta>1$,
$\epsb$ constraint is stronger for $\tan\beta\lsim 1$, and therefore these two
are the strongest and complimentary
constraints present in the charged Higgs sector of the model.

\vskip.25in
\centerline{ACKNOWLEDGEMENTS}

The author thanks Professor T.~K.~Kuo for very helpful suggestions and reading
the manuscript.
This work has been supported by the World Laboratory.


%
\def\NPB#1#2#3{Nucl. Phys. B {\bf#1} (19#2) #3}
\def\PLB#1#2#3{Phys. Lett. B {\bf#1} (19#2) #3}
\def\PLIBID#1#2#3{B {\bf#1} (19#2) #3}
\def\PRD#1#2#3{Phys. Rev. D {\bf#1} (19#2) #3}
\def\PRL#1#2#3{Phys. Rev. Lett. {\bf#1} (19#2) #3}
\def\PRT#1#2#3{Phys. Rep. {\bf#1} (19#2) #3}
\def\MODA#1#2#3{Mod. Phys. Lett. A {\bf#1} (19#2) #3}
\def\IJMP#1#2#3{Int. J. Mod. Phys. A {\bf#1} (19#2) #3}
\def\TAMU#1{Texas A \& M University preprint CTP-TAMU-#1}
\def\ARAA#1#2#3{Ann. Rev. Astron. Astrophys. {\bf#1} (19#2) #3}
\def\ARNP#1#2#3{Ann. Rev. Nucl. Part. Sci. {\bf#1} (19#2) #3}

\newpage

%
{\bf Figure Captions}
\begin{itemize}

\item Figure 1: The regions in $(m_{H^\pm},\tan\beta)$ plane excluded by
the CLEO II bound $\brbsg<5.4\times10^{-4}$ at $95\%$C.~L., for $m_t=130, 150
\GeV$ in 2HDM. The excluded regions
lie to the left of each solid curve. The excluded regions by the LEP value
$\epsb=-0.00592$ at $90\%$C.~L. lie below each dotted curve.
The values of $m_t$ used are as indicated.
\end{itemize}


\begin{thebibliography}{99}
\bibitem{assume1} We have assumed that the updated analysis would not change
the conclusions in Ref\cite{BargerLE} drastically.
\end{thebibliography}
\end{document}